\documentclass[reqno]{amsart}
\usepackage{amssymb}
\usepackage{enumerate}
\usepackage{hyperref}
\usepackage{doi,url}

\numberwithin{equation}{section}
\newtheorem{theorem}{Theorem}[section]

\newtheorem{lemma}[theorem]{Lemma}
\newtheorem{corollary}[theorem]{Corollary}
\newtheorem{definition}[theorem]{Definition}

\DeclareMathOperator{\supp}{supp}
\DeclareMathOperator{\tr}{tr}

\renewcommand\L{\mathrm{L}}
\newcommand\R{\mathbb R}
\newcommand\N{\mathbb N}

\newcommand\Z{\mathbb Z}

\newcommand\di{\mathrm d}

\newcommand\T{\mathcal{T}}

\newcommand\e{\mathrm{e}}

\renewcommand\P{\mathbb P}
\newcommand\E{\mathbb E}

\newcommand{\pr}{\prime}

\newcommand{\bom}{{\boldsymbol{{\omega}}}}

\newcommand\beq{\begin{equation}}
\newcommand\eeq{\end{equation}}
\newcommand{\qnorm}[1]{\left\lVert\!\left\Vert #1 \right\rVert\!\right\rVert}
\newcommand{\abs}[1]{\left\lvert #1 \right\rvert}
\newcommand{\norm}[1]{\left\lVert #1 \right\rVert}

\newcommand{\set}[1]{\left\{ #1 \right\}}
\newcommand{\pa}[1]{\left( #1 \right)}

\newcommand{\eq}[1]{\eqref{#1}}

\newcommand{\up}[1]{^{(#1)}}

\newcommand{\qtx}[1]{\quad\text{#1}\quad}

\begin{document}

\title[Local Wegner  estimates]{Local Wegner  and Lifshitz tails estimates for the density of states  for continuous random Schr\"odinger operators}

\author[J.-M. Combes]{Jean-Michel Combes}
\address[Combes]{Centre de Physique Th\'eorique, Aix-Marseille Universit\'e et Universit\'e du Sud Toulon  Var, CNRS UMR 6207,
F-83130 La Garde, France}
\email{combes@cpt.univ-mrs.fr}

\author[F. Germinet]{Fran\c cois Germinet}
\address[Germinet]{Universit\'e de Cergy-Pontoise,
CNRS UMR 8088,  D\'epartement de Math\'ematiques,
F-95000 Cergy-Pontoise, France}
\email{germinet@math.u-cergy.fr}

\author[A. Klein]{Abel Klein}
\address[Klein]{University of California, Irvine,
Department of Mathematics,
Irvine, CA 92697-3875,  USA}
\email{aklein@uci.edu}

\thanks{A.K. was  supported in part by the NSF under grants DMS-0457474 and DMS-1001509.}

 \subjclass[2010]{Primary 82B44; Secondary 47B80, 60H25, 81Q10}


\begin{abstract}
We introduce and prove local Wegner estimates for continuous generalized Anderson Hamiltonians, where the single-site random variables are independent but not necessarily identically distributed.  In particular, we get Wegner estimates with a constant that goes to zero as we approach the bottom of the spectrum.    As an application, we show that the (differentiated) density of states exhibits the same Lifshitz tails upper bound as the integrated density of states.
\end{abstract}

\maketitle



\section{Introduction}

In this paper we introduce and prove local Wegner estimates for continuous generalized Anderson Hamiltonians, where the single-site random variables are independent but not necessarily identically distributed.     In particular, we get Wegner estimates with a constant that goes to zero as we approach the bottom of the spectrum.  As an application of local Wegner estimates, we show that the (differentiated) density of states exhibits the same Lifshitz tails upper bound as the integrated density of states.

 We consider continuous generalized Anderson Hamiltonians, which are  random Schr\"odinger 
operators on 
$\mathrm{L}^2(\mathbb{R}^d)$ of the type
\beq\label{AndH}
H_{\bom}: =  -\Delta + V_{\mathrm{per}} +
V_{\bom} ,
\eeq
where: $\Delta$ is the $d$-dimensional Laplacian operator; $V_{\mathrm{per}}$ is a bounded $q \Z^d$-periodic potential with $q\in \N$; and  $V_{\bom}$ is an alloy-type random potential:
\beq
V_{\bom} (x):= 
\sum_{j \in \Z^d} \omega_j \,  u_j(x), \quad \text{with } \quad u_j(x )=u(x-j) ,\label{AndV}
\eeq
where   
the single site potential $u$ is a  nonnegative bounded 
measurable function
on $\R^{d}$ with compact support, uniformly 
bounded away from zero in
a neighborhood of the origin, and
$\bom=\{ \omega_j \}_{j\in
\Z^d}$ is a family of independent (not necessarily identical)
 random
variables, such that, with  $\mu_j$ denoting the  probability 
distribution  of $\omega_j$,  
\beq
\bigcup_{j\in \Z^d} \supp \mu_j \subset [M_-,M_+]  \qtx{for  some }  \infty < M_- < M_+<\infty.
\eeq
Without loss of generality we specify $\norm{u}_\infty=1$,  which can be aways be achieved by rescaling the $\mu_j$.  In this paper we assume 
  that  $\mu_j$ has no atoms (i.e., $\mu_j$is a continuous measure)  for all  $j\in \Z^d$.
 The (ergodic) Anderson Hamiltonian is the special case when
the $\{ \omega_j \}_{j\in\Z^d}$ are  identically distributed, i.e., $\mu_j=\mu$ for all $j\in \Z^d$.

 Given a finite Borel  measure $\nu$ on $\R$ and $s\ge 0$, we let $S_\nu(s):=\protect{ \sup_{a \in \R}
\nu([a,a+s])}$, 
  the concentration function of $\nu$, and set 
   \beq \label{defQ}
Q_\nu(s):=
\left\{ 
\begin{array}{ll}
\norm{\rho}_\infty s &\mbox{if $\nu$ has a bounded density $\rho$} 
\\
 8 S_\nu(s) &\mbox{otherwise}
\end{array}.
\right. 
\eeq
  $Q_\nu(s)$ is continuous on $[0,\infty[$ if and only if
 the measure $\nu$ has no atoms, in which case $\lim_{s
\downarrow 0} Q_\nu(s)=Q_\nu(0)=0$ \cite{HT}.

The finite volume operator $H_{\bom}\up{\Lambda}$,  the restriction   of $H_{\bom}$ to  a  finite box $\Lambda$ with periodic boundary condition,   has a finite number of eigenvalues in a given bounded interval $I\in \R$.
Fluctuations of these eigenvalues due to the  random variables $\{ \omega_j \}_{j\in
\Z^d}$ play a crucial role in the understanding of the localization properties of $H_\bom$. When averaging over a single random variable, the fluctuations of the eigenvalues are controlled thanks to a spectral averaging principle:   given a trace class operator $S\ge 0$, we have \cite{CH,CHK2}
\begin{equation}\label{sa0}
\E_{\omega_j}\set{ \tr \set{\sqrt{u_j}\chi_I(H_{\bom}\up{\Lambda}) \sqrt{u_j}S}} \le \pa{\tr S} Q_{\mu_j}(\abs{I}) \qtx{for} j \in \Z^d\cap \Lambda.
\end{equation}
 Averaging over all the random variables, the expectation of the number of eigenvalues falling in an interval $I$ is controlled thanks to the celebrated Wegner estimate \cite{W,CH,CHK1,CHK2,Kle2}:  
\begin{equation}\label{Wegner0}
\E\set{ \tr \chi_I(H_{\bom}\up{\Lambda}) } \le K_W Q_{\Lambda}(\abs{I}) \abs{\Lambda},
\end{equation}
where
\beq\label{QLambda}
Q_{\Lambda}(s):= \max_{j \in \Lambda \cap \Z^d } Q_{\mu_j}(s) ,\eeq
and the constant $K_W$ depends on the parameters $d, u, M_\pm$, and $\sup I$.

An estimate of the form
\begin{equation} \label{locWegner0}
\max_{j \in \Lambda \cap \Z^d} \E \set{\tr \chi_I(H_{\bom}\up{\Lambda}) u_j} \le K_{LW}Q_{\Lambda}(\abs{I}) 
\end{equation}
will be called a local Wegner estimate.  If the
  generalized Anderson Hamiltonian $H_{\bom}$ satisfies the  covering condition
 $\sum_{j\in \Z^d\cap \Lambda} u_j \ge C \chi_\Lambda>0$ with $C>0$, the Wegner estimate \eq{Wegner0} can be immediately derived  from the local Wegner estimate   \eqref{locWegner0}. 
If the random variables  $ \{ \omega_j \}_{j\in
\Z^d}$ are identically distributed, under the above covering condition  it is equivalent to investigate local and global Wegner estimates. Indeed, using the covariance property of the model, in this case  there exist constants $C_1$ and $C_2$ so that for any $j\in\Lambda$ we have 
\begin{equation}
\tfrac{C_1}{|\Lambda|}\E \set{\tr \chi_I(H_{\bom}\up{\Lambda})} \le \E \set{\tr \chi_I(H_{\bom}\up{\Lambda}) u_j}
\le \tfrac{C_2}{|\Lambda|}\E \set{\tr \chi_I(H_{\bom}\up{\Lambda}) }.
\end{equation}

In Theorem~\ref{thmlocalWeg} we prove local Wegner estimates for generalized Anderson Hamiltonians with a covering condition.  Moreover, we provide bounds on the local Wegner constant $K_{LW}$ in \eqref{locWegner0} that vanish as the energy approaches the bottom of the spectrum. These results are new, both in the ergodic and non-ergodic cases, and yield bounds on the density of states. Theorem~\ref{thmlocalWeg} extends \cite[Lemma~4.1]{CGK2} to more general single site probability distributions and to a less restrictive covering condition.

Using a local Wegner estimate, we prove in Corollary~\ref{thmLT}  that the differentiated density of states exhibits the same Lifshitz tails upper bound as the integrated density of states  for Anderson Hamiltonians with a covering condition and a single-site probability distribution with a bounded density.    This result had been shown to hold for  discrete Anderson models \cite{CGK3}.

\section{Results}

 We write
\beq \label{box}
\Lambda_{L}(x):= x +\left[-\tfrac L 2, \tfrac L 2\right[^d
\eeq
for the  (half open-half closed)
box of side $L>0$ centered at $x\in \R^d$.  By $\Lambda_L$ we denote a box $\Lambda_{L}(x)$ for some $x \in \R^d$.  Given a box $\Lambda=\Lambda_{L}(x)$, we 
set $\widetilde{\Lambda}= \Lambda \cap \Z^d$. If $B$ is a set, we write  $\chi_B$ for its characteristic function. 
 We set $\chi^{(L)}_x:=\chi_{\Lambda_L(x)}$, with  $\chi_x:=\chi^{(1)}_x$  The  Lebesgue measure of a  Borel set $B \subset \R$ will be denoted  by $\abs{B}$.  
  By a constant we will always mean a finite constant. Constants such as $C_{a,b,\ldots}$  will be finite and depending only on the parameters or quantities $a,b,\ldots$; they will be independent of other  parameters or quantities in the equation.  Note that $C_{a,b,\ldots}$ may stand for different constants in different sides of the same inequality.

Before stating our results,
we  normalize  a generalized Anderson Hamiltonian  $H_{\bom}$ as follows.   We  first require  $\inf_{j \in \Z^d} \inf \supp \mu_j = 0$, which  can always be realized by 
changing   the periodic potential  $V_{\mathrm{per}}$. 
We then adjust    $V_{\mathrm{per}}$ by adding a constant so $\inf \sigma\pa{ -\Delta + V_{\mathrm{per}}}= 0$, in which case $[0, E_*] \subset\sigma\pa{ -\Delta + V_{\mathrm{per}}}$  for some $E_*>0$.  The result is a normalized  generalized Anderson Hamiltonian  as in the following definition, equal to the original  generalized Anderson Hamiltonian  given in \eq{AndH}-\eq{AndV} plus a nonrandom constant. We also assume that  the single site probability distributions have no atoms.

 \begin{definition} \label{defAndH}
 A normalized  generalized Anderson Hamiltonian is  a  generalized  Anderson Hamiltonian $H_{\bom}$ as in \eq{AndH}--\eq{AndV}, such that:
 \begin{enumerate} 
\item The free Hamiltonian $ H_0 := -\Delta + V_{\mathrm{per}}$ has $0$ as the bottom of its spectrum:
\beq \label{AndH287}
\inf \sigma(H_0)= 0 .
\eeq

\item The single site potential $u$ is a measurable function on $\R^d$
 with 
 \begin{equation} \label{u}
\norm{u}_\infty=1 \quad \text{and} \quad u_{-}\chi_{\Lambda_{\delta_{-}}(0)}\le u \le \chi_{\Lambda_{\delta_{+}}(0)}, \qtx{where} u_{-}, \delta_{\pm}\in ]0,\infty[ ;
\end{equation}
we set
\beq
U_+:= \norm{\textstyle{\sum}_{j \in \Z^d} \,  u_j}_\infty\le \max\set {1, \delta_+^d}.   \label{U+}
\eeq

\item  $\bom=\{ \omega_j\}_{j \in \Z^d}$ is a family of independent random variables,  such that for
all  $j\in \Z^d$ the probability 
distribution $\mu_j$ of $\omega_j$ has no atoms and
\beq\label{suppmuM}
0=  \inf_{j\in \Z^d} \inf\supp \mu_j < M:=  \sup_{j\in \Z^d} \sup \supp \mu_j<\infty .
\eeq

\end{enumerate}
$H_\bom$ is a normalized Anderson Hamiltonian 
if the $\{ \omega_j \}_{j\in\Z^d}$ are  identically distributed, i.e., $\mu_j=\mu$ for all $j\in \Z^d$. In this case $\mu$ is a probability measure with no atoms such that
 \beq \label{mu00}
\set{0  ,M}\subset \supp \mu \subset   [0,M],\quad \text{where} \quad  M\in ]0,\infty[. 
\eeq
 \end{definition}

Without loss of generality, we will always assume that a generalized Anderson Hamiltonian $H_\bom$   is a normalized generalized Anderson Hamiltonian.    In particular, Anderson Hamiltonians will also be understood to be normalized.

We will  need generalized Anderson Hamiltonians with more structure.  We set
\beq
\Gamma(j_0,K):= j_0 +  K\Z^d, \qtx{where} j_0 \in \Z^d \qtx{and} K\in \N.
\eeq
Note that for any $j\in \Z^d$  there exists $j^\pr\in \Gamma(j_0,K)$ such that $j \notin  \Gamma(j^\pr,2K)\subset \Gamma(j_0,K)$.

 \begin{definition} A generalized Anderson Hamiltonian $H_\bom$ has a spine if there exist $j_0\in \Z^d$ and $K \in \N$ such that the random variables  $\{ \omega_j\}_{j \in \Gamma(j_0,K)}$ are identically distributed.  In this case we will call $\Gamma=\Gamma(j_0,K)$  a spine of order $K$ for  $H_\bom$ and set $\mu_{\Gamma} := {\mu_{j}} $ for $j\in \Gamma$.
  \end{definition}
 
An  Anderson Hamiltonian $H_{\bom}$ (in this language  a generalized Anderson Hamiltonian with a spine of order $1$)      is a $q\Z^d$-ergodic family of
random self-adjoint operators.
It follows from standard results (cf. \cite{KiM})
that there exists fixed subsets $\Sigma$,  $\Sigma_{\mathrm{pp}}$, $\Sigma_{\mathrm{ac}}$  and $\Sigma_{\mathrm{sc}}$ of $\R$ so that the spectrum $\sigma(H_{\bom})$
of $H_{\bom}$,  as well as its pure point, 
absolutely continuous, and singular continuous  components,
are equal to these fixed sets with probability one.  With our normalization, 
the non-random spectrum $\Sigma$ of    an  Anderson Hamitonian  $H_{\bom}$
satisfies (cf. \cite{KiM2})
\beq
\sigma\pa{ H_0} \subset \Sigma \subset [0, \infty[,
\eeq
with $\inf \Sigma=0 $ and $ [0, E_*] \subset\Sigma$ for some $E_*=E_*(V_{\mathrm{per}})>0$.
Note that $ \Sigma= \sigma\pa{ -\Delta } =[0,\infty[  \quad \text{if} \quad  V_{\mathrm{per}}=0$.

A generalized Anderson Hamiltonian $H_\bom$ is not, in general, an ergodic family of
random self-adjoint operators, so the above considerations do not apply, and its spectrum  is a random set.  But it follows from Definition~\ref{defAndH} that
\beq
\sigma(H_\bom) \subset [0,\infty[ \quad \text{with probability one}.
\eeq
Note furthermore that if the  generalized Anderson Hamiltonian $H_\bom$ has a spine $\Gamma$ of order $K$, then  \beq
H_{\bom_{\Gamma}}= H_{0}+ V_{\bom_{\Gamma}} \qtx{where} \bom_{\Gamma}=\set{\omega_{j}}_{j\in \Gamma} \qtx{and} V_{\bom_{\Gamma}} (x):= 
\sum_{j \in \Gamma} \omega_j \,  u_j(x), \label{AndGamma}
\eeq
is a $qK\Z^{d}$-ergodic family of random self-adjoint operators, and the above considerations for Anderson Hamiltonians apply. ($H_{\bom_{\Gamma}}$  is exactly like an Anderson Hamiltonian, except that  the single site potentials are located in $K\Z^{d}$ instead  of $\Z^{d}$.)

Let  $H_{\bom}$ be a generalized Anderson Hamiltonian.  Finite volume operators are defined for 
finite boxes $\Lambda=\Lambda_L(j_0)$, where $j_0\in \Z^d$ and $L \in 2q\N$,  $ L > \delta_+$.   Given   such $\Lambda$, we will consider the random Schr\"odinger operator $H_\bom^{(\Lambda)}$ on $\L^2(\Lambda)$ given by the restriction of the generalized Anderson Hamiltonian  $H_\bom$ to $\Lambda$ with periodic boundary condition. To do so, we identify $\Lambda$ with a torus
in the usual way by identifying opposite edges, and  define finite volume operators  
\begin{align}\label{finvolH}
H_{\bom}\up{\Lambda} :=H_{0}\up{\Lambda}+ V_{\bom}\up{\Lambda} \quad \text{on}   \quad \L^{2}(\Lambda).
\end{align}
The finite volume free Hamiltonian $H_{0}\up{\Lambda}$ is given by
\beq
H_{0}\up{\Lambda}:= - \Delta\up{\Lambda} +  V_{\mathrm{per}}\up{\Lambda} \quad \text{on}   \quad \L^{2}(\Lambda),
\eeq
where 
$\Delta\up{\Lambda}$ is the  Laplacian on $\Lambda$ with periodic boundary condition and   
$V_{\mathrm{per}}\up{\Lambda}$ is the restriction of $ V_{\mathrm{per}}$  to $\Lambda$.
The random potential $V_{\bom}\up{\Lambda}$ is the restriction of $V_{\bom\up{\Lambda}}$ to $\Lambda$, where, given $\bom=\set{\omega_i}_{i \in \Z^d}$ , we define $\bom\up{\Lambda}=\set{\omega\up{\Lambda}_i}_{i \in \Z^d}$ by
\beq\begin{split}
\omega\up{\Lambda}_i =\omega_i \quad \text{if}  \quad i \in \widetilde{\Lambda},  \quad
\omega\up{\Lambda}_i =\omega\up{\Lambda}_k  \quad \text{if}  \quad k-i \in L\Z^d .
\end{split}\eeq
Note that the random finite volume operator $ H_{\bom}\up{\Lambda}$ is not covariant with respect to translations in the torus  unless $ H_{\bom}$ is an Anderson Hamiltonian.

Given $j \in \widetilde{\Lambda}$, we set 
\beq\label{ujLa}
u_j\up{\Lambda}(x):= \sum_{k \in j + L\Z^d} u_k(x) \qtx{and} \chi_j\up{\Lambda}(x):= \sum_{k \in j + L\Z^d} \chi_k(x) \qtx{for} x\in \Lambda,
\eeq
and rewrite $ V_{\bom}\up{\Lambda}$ as
\beq
 V_{\bom}\up{\Lambda}= \sum_{j \in \widetilde{\Lambda}} \omega_j u_j\up{\Lambda}.
\eeq
We will often  abuse the notation and just write $u_j$ and $\chi_j$ instead of $u_j\up{\Lambda}$ and $  \chi_j\up{\Lambda} $ when working with finite volume operators.  Note that
\beq\label{covLambdachi}
 \sum_{j \in \widetilde{\Lambda}}  \chi_j\up{\Lambda}(x)  =1   \qtx{for all} x \in \Lambda.
\eeq
When the covering condition $\delta_-\ge 1$ (see \eqref{u}) holds,  we have
\beq\label{covLambda}
 \sum_{j \in \widetilde{\Lambda}}  u\up{\Lambda}_j(x)  \ge  u_- \qtx{for all} x \in \Lambda.
\eeq

Given  a finite Borel measure $\nu$ on $\R$ with no atoms and   finite moments, and $m\ge 1$, we set (recall  \eq{QLambda})
\beq
Q_\nu\up{m}(s):=Q_{\nu\up{m}}(s), \qtx{where}  \di \nu\up{m}(t)= (1 + \abs{t}^m) \di \nu(t).
\eeq
 In particular, if $\supp \nu \subset [0,M]$  (cf.  \eq{suppmuM}) we have
\beq
Q_\nu\up{m}(s)\le \pa{1 +M}^m Q_{\nu}(s) \qtx{for} m \ge 1.
 \eeq

The finite Borel  measure $\nu$  is said to be H\" older continuous of order  $\alpha \in ]0,1]$ if there exists a  constant $C_{\nu,\alpha}$ such that
\beq\label{Holdercts}
Q_\nu(s) \le C_{\nu,\alpha} s^\alpha \qtx{for all} s \in [0,1] .
\eeq
If in addition  $\supp \nu \subset [0,M]$, it follows that $\nu\up{m}$  is also  H\" older continuous of order  $\alpha$ for all $m\ge 1$:
\beq\label{Holderctsm}
Q_\nu\up{m}(s)\le  C_{\nu,\alpha,m}   s^\alpha \qtx{with}   C_{\nu,\alpha,m}\le  C_{\nu,\alpha} \pa{1 +M}^m.
 \eeq
If $\nu$ has a bounded density $\rho$  (i.e., $\alpha=1$) and $\supp \nu \subset [0,M]$,  then 
\eq{Holdercts} holds with
$C_{\nu,1}=\norm{\rho}_\infty$.  In this case,   for all $m\ge 1$ the measure  $\nu\up{m}$ has a bounded density $\rho\up{m}(t)= (1 + {t}^m)\rho(t) $, and 
\beq\label{Holderctsmrho}
Q_\nu\up{m}(s)\le \norm{\rho\up{m}}_\infty   s \qtx{with}  \norm{\rho\up{m}}_\infty \le  \pa{1 +M}^m \norm{\rho}_\infty.
 \eeq
 
Let  $H_{\bom}$ be a generalized Anderson Hamiltonian.   
If $B \subset \R$ is a Borel set, we write $P_\bom^{(\Lambda)}(B):=\chi_B\big(H_\bom^{(\Lambda)}\big)$ and $P_\bom(B):=\chi_B(H_\bom)$ for the spectral projections. 
 Let $E_0 >0$, $I \subset [0,E_0]$ an interval, and   consider a box $\Lambda=\Lambda_L(j_0)
$, where  $L \in 2q\N$,  $ L > \delta_+$, and $j_0 \in \Z^d$.     If $H_{\bom}$ satisfies the covering condition $\delta_-\ge 1$ (see \eqref{u}), we have the Wegner  estimate \cite{CH,CHK2,CGK2} (see also \eq{localWegreal} below) 
\beq\label{Wegcov}
\E   \set{\tr P_\bom^{(\Lambda)}(I)} \le K_{W}(E_0) Q_{\Lambda}(\abs{I}) \abs{\Lambda}.
\eeq
  Without assuming the covering condition, a careful reading of  \cite{CHK2}, as in \cite[Appendix~B]{GKM}, gives
\beq\label{Wegnocov}
\E   \set{\tr P_\bom^{(\Lambda)}(I)} \le K_{W}(E_0) Q\up{m_d}_{\Lambda}(\abs{I}) \abs{\Lambda} \qtx{with} m_d=2^{2 + \frac {\log d}{\log 2}},
 \eeq
where
 \beq \label{WegnocovM}
 Q_{\Lambda}\up{m}(s):= \max_{j \in \widetilde{\Lambda} } Q_{\mu_j}\up{m}(s)\le (1 + M)^{m}Q_{\Lambda}(\abs{I}) \qtx{for} m\ge 1.
 \eeq 
  The constants $ K_{W}(E_0)$ in \eq{Wegcov}  and \eq{Wegnocov}  depend only on  $d, V_{\mathrm{per}},\delta_+,u_-$; they do not depend on the probability distributions $\mu_j$.
   If the  generalized Anderson Hamiltonian $H_\bom$ has a spine $\Gamma$ we set  $Q_\Gamma\up{m}=Q_{\mu_\Gamma}\up{m}$.

We now state our  local Wegner estimates. We set  $[[\frac d 4]]= \lfloor \frac d 4\rfloor  +1$,  the  smallest integer $> \frac d 4$.

\begin{theorem}\label{thmlocalWeg}
Let $H_{\bom}$ be a generalized Anderson Hamiltonian with $\delta_-\ge 1$, and let $\Lambda=\Lambda_L(j_0)
$, where  $L \in 2q\N$,  $ L > \delta_+$, and $j_0 \in \Z^d$.
\begin{enumerate}
\item 
Given $E_0>0$, for all intervals $I \subset [0,E_0]$ we have
\beq\label{localWeg1}
\max_{j \in \widetilde{\Lambda}} \E   \set{\tr P_\bom^{(\Lambda)}(I) u\up{\Lambda}_{j}}\le  C_{d,\norm{ V_{\mathrm{per}}^-},\delta_+}  u_-^{-\frac 3 2} (1 + E_0)^{2[[\frac d 4]]}(1 +\log (1 + E_0))Q_\Lambda(\abs{I}),
\eeq
yielding the Wegner estimate
\beq\label{localWegreal}
\E   \set{\tr P_\bom^{(\Lambda)}(I)}  \le  C_{d,\norm{ V_{\mathrm{per}}^-},\delta_+} u_-^{-\frac 5 2}(1 + E_0)^{2[[\frac d 4]]}(1 +\log (1 + E_0))Q_\Lambda(\abs{I})\abs{\Lambda}.
\eeq

\item  Suppose the generalized Anderson Hamiltonian $H_\bom$ has a spine $\Gamma$ of order $K$. Given $\eta \in ]0,\frac d 2[$,   there exists $E_1=E_1(\eta,d, V_{\mathrm{per}},\delta_\pm,u_-,\mu_\Gamma,K)>0$, such that for all  $E_0 \in ]0,E_1[$   and intervals  $I \subset [0,E_0]$,  we have
\beq\label{localWeg344}
\max_{j \in \widetilde{\Lambda}} \E   \set{\tr P_\bom^{(\Lambda)}(I) u\up{\Lambda}_{j}}\le \e^{-E_0^{-\frac d2+\eta}}Q_{\Lambda}(\abs{I}),
\eeq
yielding the Wegner estimate
\beq\label{localWegreal344}
\E   \set{\tr P_\bom^{(\Lambda)}(I)}  \le  u_-^{-1}\e^{-E_0^{-\frac d2+\eta}}Q_\Lambda(\abs{I})\abs{\Lambda},
\eeq
for $L$ large (how large depending on $E_0, d, V_{\mathrm{per}},\delta_\pm,u_-,\mu_\Gamma,K,\eta$).  

\item  Suppose the generalized Anderson Hamiltonian $H_\bom$ has a spine $\Gamma$ of order $K$ with a  H\" older continuous single-site probability distribution $\mu_\Gamma$ of order $\alpha$, and let $L \in 2qK\N$.  Then there exists $E_1=E_1(d, V_{\mathrm{per}},\delta_\pm,u_-,\mu_\Gamma,K)>0$, such that for all  $E_0 \in ]0,E_1[$, intervals  $I \subset [0,E_0]$, and $\eta \in ]0,1[$,  we have
\begin{align}\label{localWeg3}
\max_{j \in \widetilde{\Lambda}} \E   \set{\tr P_\bom^{(\Lambda)}(I) u\up{\Lambda}_{j}} &\le    C_\eta\pa{ C_{\mu_\Gamma} \pa{ 2 \alpha E_0  \log \tfrac 1{2\alpha E_0 C_{\mu_\Gamma}^{\frac 1 \alpha}}}^\alpha}^{1-\eta}  Q_{\Lambda}(\abs{I})\\
&\le C_{\eta,\mu_\Gamma} E_0^{\alpha(1-\frac \eta 2)}  Q_{\Lambda}(\abs{I}),\label{localWeg3449}
\end{align}
yielding the Wegner estimate
\beq\label{localWegreal3449}
\E   \set{\tr P_\bom^{(\Lambda)}(I)}  \le   C_{\eta,\mu_\Gamma} u_-^{-1}E_0^{\alpha(1-\frac \eta 2)}Q_\Lambda(\abs{I})\abs{\Lambda},
\eeq
for $L$ large (how large depending on $d, V_{\mathrm{per}},\delta_\pm,u_-,\mu_\Gamma,K$), where   $C_\eta=C_{d, V_{\mathrm{per}},\delta_\pm,u_-, K,E_1,\eta}$, $C_{\mu_\Gamma}=C_{\mu_\Gamma,\alpha,m_d}$ as in \eq{Holderctsm}  with $m_d$ given in  \eq{Wegnocov},
and $C_{\eta,\mu_\Gamma}= C_{d, V_{\mathrm{per}},\delta_\pm,u_-, K,E_1,\mu_\Gamma,\eta}$.
\end{enumerate}
\end{theorem}

Part (i), namely \eq{localWeg1}, gives a local version of of the Wegner estimates \eq{Wegcov} and \eq{WegnocovM}.  It is of the form given in \eq{locWegner0}, valid at all energies $E_0$ with a constant $K_{LW}=K_{LW}(E_0)$, but the constant does not get small as $E_0 \downarrow 0$.  
Parts (ii) and (iii) provide  local Wegner estimates valid for small $E_0$ with $\lim_{E_0 \downarrow 0}K_{LW}(E_0)=0$..  Part (ii) requires less hypotheses, and seems to provide a stronger result.  But we believe that the energy interval $[0,E_1]$ where the estimates hold is bigger in (iii).  The proof of (ii) takes advantage of the Lifshitz tails estimate, and is thus valid in an energy interval at the bottom of the spectrum where we have Lifshitz tails.  The proof of (iii) uses  dynamical localization estimates, and is  valid in   the energy interval where we can perform the bootstrap multiscale analysis of \cite{GKboot}, which in principle is larger than the region of Lifshitz tails. In addition, \eq{localWeg3}, unlike \eq{localWeg344},  shows the explicit dependence of the constant on the single-site probability distribution $\mu_\Gamma$. (This is the reason why we state \eq{localWeg3} in addition to \eq{localWeg3449}.)  Note that when $\mu_\Gamma$ has a bounded density $\rho$,  we have (recall \eq{Holderctsmrho}) $C_{\mu_\Gamma}=  \norm{\rho\up{m_d}}_\infty \le  \pa{1 +M}^{m_d} \norm{\rho}_\infty$.

The Wegner estimates \eq{localWegreal344} and \eq{localWegreal3449}, with constants that go to zero as $E_0 \to 0$,  only require the  covering condition  $\delta_- \ge1$.  This is a substantial improvement on the similar  Wegner estimate given in \cite[Lemma~4.1(i)]{CGK2}, which requires the double covering condition $\delta_- \ge2$.

An  Anderson Hamiltonian $H_\bom$ satisfies a Lifshitz tails estimate, which asserts that its integrated density of states  $N(E)$ has  exponential fall off as the energy $E$ approaches the bottom of the spectrum. The finite volume operator $H_\bom^{(\Lambda)}$ has a compact resolvent, and hence its ($\bom$-dependent) spectrum consists of isolated  eigenvalues with finite multiplicity. 
 We recall  that the integrated density of states (IDS)  for $H_\bom$ is given, for a.e.  $E \in \R$, by ($\Lambda_L=\Lambda_L(0)$)
\beq \label{N(E)}
N(E):= \lim_{L \to \infty} \abs{\Lambda_L}^{-1} \tr  \, \chi_{]- \infty,E]}\pa{H_{\bom}\up{\Lambda_L}}\quad \text{for $\P$-a.e. $\bom$},
\eeq
in the sense that the limit exists and is the same  for $\P$-a.e. $\bom$   (cf.\  \cite{CL,N,PF}). 
Recalling that with our normalization the bottom of the spectrum is at $0$,  the IDS  satisfies the Lifshitz tails estimate (e.g., \cite[Corollary~2.2 and Remark~7.1]{Klop99})
\beq\label{LT}
\lim_{E \downarrow 0}     \frac{\log \abs{\log    N(E)}}{\log E}   
\le - \frac d2 .
\eeq
Equality is actually known to hold in \eqref{LT}.

Since the integrated density of states  $N(E)$ is an increasing function, it has a derivative $n(E):= N^{\pr}(E)\ge 0$  almost everywhere,  the density of states. Note that by ergodicity with respect to $q\Z^d$ we have 
\beq\label{NEq}
N(E)=q^{-d}\E \set{\tr \chi_0^{(q)} \chi_{]-\infty,E]}(H_{\bom})\chi_0^{(q)}},
\eeq
 and hence
\beq
N(E^\pr)-N(E)\le  q^{-d} \E \set{\tr \chi_0^{(q)} \chi_{]E,E^\pr]}(H_{\bom})\chi_0^{(q)}}\qtx{for} E \le E^\pr.
\eeq
As a consequence,  if  the single-site probability distribution $\mu$ has a bounded density $\rho$, and  the local Wegner estimate \eqref{locWegner0} holds for intervals $I \subset [0,E_0]$, we conclude that  
\beq\label{preLT}
n(E)\le  q^{-d} K_{LW} \norm{\rho}_\infty \qtx{for a.e.}  E \in  [0,E_0].
\eeq    

The following corollary, which provides an exponentially small bound for the density of states within the regime of Lifshitz tails,  is an immediate corollary of Theorem~\ref{thmlocalWeg}(ii), using \eqref{localWeg344} and \eq{preLT}.

\begin{corollary}\label{thmLT}
Let $H_{\bom}$ be an  Anderson Hamiltonian with $\delta_-\ge 1$, whose  single-site probability distribution $\mu$ has a bounded density $\rho$.  Then there exists a Borel set $\mathcal{N}\subset [0,1]$ of zero Lebesgue measure such that
\beq\label{LTdos}
\lim_{E \downarrow 0; \ E \notin \mathcal{N}}     \frac{\log \abs{\log    n(E)}}{\log E}   
\le - \frac d2 .
\eeq
\end{corollary}

The same Lifshitz tails estimate for the density of states holds for the discrete Anderson  model \cite{CGK3}.

\section{Proof of local Wegner estimates}

\subsection{A simple Lemma}

\begin{lemma}\label{lemobs}
Let $H=H_0+W$, where $H ,H_0$ are semi-bounded  self-adjoint operators, say $H,H_{0}\ge -\Theta$ for some $\Theta>0$,  such that    $\pa{H+ \Theta+1}^{-p}$
is a trace class operator for some $p >0$, and $W$ is a  bounded  self-adjoint operator. Let $E_0 \in \R$.  Let $f,h$ be bounded Borel measurable nonnegative functions with compact support such that
$f= \chi_{(-\infty, E_0]} f$,  $h= \chi_{[ E_0,\infty)} h $,  and $H_{0}h(H_0)$ is a bounded operator. Then $ f(H)W h(H_0)$ is trace class and
\begin{equation}\label{neg}
\tr f(H)Wh(H_0) \le 0 .
\end{equation}
In particular, if $f,g$ are  bounded Borel measurable nonnegative functions such that
$f= \chi_{(-\infty, E_0]} f$ and $\chi_{(-\infty, E_0]}\le  g\le 1$, we have $ f(H)W$ and $ f(H) W g(H_0)$ trace class, and 
 \begin{equation}
\tr f(H)W \le \tr f(H) W g(H_0).
\end{equation}
\end{lemma}

Note that $W$ does not need to be positive.

\begin{proof}
Let $f,h$ be as above, note that $f(H)$ is trace class. Then, as  $W = H -H_0$, we have
 \begin{align}
 \tr f(H)W h(H_0)= \tr f(H)H h(H_0)- \tr f(H)H_0  h(H_0),
 \end{align}
 where both $ f(H)H h(H_0)$ and $ f(H)H_0  h(H_0)$ are trace class operators. Moreover,
 \begin{align}
  \tr f(H)H h(H_0)& \le E_0  \tr f(H) h(H_0),\\
  \tr f(H)H_0  h(H_0) &\ge E_0  \tr f(H) h(H_0),
 \end{align}
 so \eq{neg} follows. 
 
 Now let $f,g$ be as above.  Let also $\chi_n=\chi_{(-\infty,n]}$.  Then,  using \eq{neg},
 \begin{align}\notag
 \tr f(H)W&  =\lim_{n\to \infty}  \tr f(H)W \chi_n(H_0)\\
 &  = \tr f(H) W g(H_0) + \lim_{n\to \infty}  \tr f(H)W \chi_n(H_0)\pa{1 - g(H_0)}\\
 & \le \tr f(H) W g(H_0).\notag  \qedhere
 \end{align}
\end{proof}

\subsection{Norms on random operators}\label{aprandomop}
 Given $p \in [1,\infty)$,
$\T_p$ will denote  the Banach space of bounded operators $S$
on $\mathrm{L}^2(\mathbb{R}^d, {\mathrm{d}}x)$ with
$\| S \|_{\T_p}=\| S \|_p := \left(\tr |S|^p\right)^{\frac 1p} < \infty$. 
A random operator $S_\omega$ is a strongly measurable 
map from the probability
space $(\Omega,\P)$ to bounded operators on 
$\mathrm{L}^2(\mathbb{R}^d, {\mathrm{d}}x)$.   Given $p \in [1,\infty)$,
we set
\begin{equation}
\qnorm{S_\omega}_p:=
\left\{ \E \left\{  \| S_\omega \|_p^p   \right\}\right\}^{\frac 1p}= 
\left\lVert  \| S_\omega \|_{\T_p} \right\rVert_{\text{L}^p(\Omega,\P) },
\end{equation}
and
\begin{equation}
\qnorm{S_\omega}_\infty:=
\left\lVert  \| S_\omega \| \right\rVert_{\text{L}^\infty(\Omega,\P) }.
\end{equation}
These are norms on random operators, note that
\begin{equation}\label{qcomp}
\qnorm{S_\omega}_q \le \qnorm{S_\omega}_\infty^{\frac{q-p} q} \qnorm{S_\omega}_p^{\frac pq}
\quad \text{for $1\le p\le q <\infty$},
\end{equation}
and they satisfy Holder's inequality:
\begin{equation}\label{Holdersineq}
\qnorm{S_\omega T_\omega}_r\le  \qnorm{S_\omega}_p \qnorm{T_\omega}_q
\quad \text{for $r,p,q \in [1,\infty]$ with $\tfrac 1 r= \tfrac 1 p + \tfrac 1q$}.
\end{equation}

\subsection{Proof of Theorem~\ref{thmlocalWeg}}

\begin{proof} Let $H_{\bom}$ be a generalized Anderson Hamiltonian satisfying  the covering condition $\delta_-\ge 1$ (see \eqref{u}), and  consider a box $\Lambda=\Lambda_L$, where  $L \in 2q\N$,  $ L > \delta_+$.  Let  $E_0 >0$,  $I \subset [0,E_0]$ an interval.   Let $g$ be a  bounded Borel measurable  function such that
 $\chi_{(-\infty, E_0]}\le  g\le 1$.   Given $j \in \Z^d$, we let  $\bom_j^\perp=\set{\omega_k}_{k \in \Z^d\setminus\set{j}}$, write $\bom=(\bom_j^\perp,\omega_j)$, and consider the random Schr\" odinger operator $ H_{\bom_j^\perp}=H_{\bom}- \omega_{j }u_{j}$.

To simplify the notation, we will write $ u_k$ and $\chi_k$  for  $u_k^{(\Lambda)}$ and $\chi_k^{(\Lambda)}$, and set 
\begin{align}
\hat{\chi}_k&: = u_k^{-\frac 1 2} \chi_k\le u_-^{-\frac 1 2}  \chi_k \qtx{for}  k\in
 \widetilde{\Lambda} \quad \text{(recall \eq{u} and  $\delta_- \ge 1$)},\\
P&=P^{(\Lambda)}_{\bom}(I) :=\chi_{I}(H^{(\Lambda)}_\bom),\\
\widetilde{P}_j &=\widetilde{P}^{(\Lambda)}_{j,\bom_j^\perp}(I): =g(H^{(\Lambda)}_{\bom_j^\perp}), \qtx{where} j \in \widetilde{\Lambda} \qtx{and}  H^{(\Lambda)}_{\bom_j^\perp}=H^{(\Lambda)}_{\bom}- \omega_{j }u_{j}. \label{deftildeP} 
\end{align}

 Given $j \in \widetilde\Lambda$, it follows from Lemma~\ref{lemobs}, using \eq{covLambdachi}, that 
\begin{align}
\tr P u_j & \le \tr P u_j \widetilde{P}_j  = \sum_{k \in \widetilde\Lambda} \tr P u_j \widetilde{P}_j \chi_k = \sum_{k \in \widetilde\Lambda} \tr u_k^{\frac 1 2} P u_j^{\frac 1 2} u_j^{\frac 1 2}\widetilde{P}_j \hat{\chi}_k \\
&  = \sum_{k \in\widetilde\Lambda} \tr u_k^{\frac 1 2} P u_j^{\frac 1 2} T_{j,k},\notag
\end{align}
where
\beq
T_{j,k}= u_j^{\frac 1 2}\widetilde{P}_j \hat{\chi}_k.
\eeq
It follows that 
\begin{align}
& \E \set{\tr P u_j}=\qnorm{P u_j^{\frac 12}}_{2}^{2}\le \sum_{k \in\widetilde\Lambda}\qnorm{u_k^{\frac 1 2} P u_j^{\frac 1 2} T_{j,k}}_{1}\\ \notag
& \quad \le \sum_{k \in \widetilde\Lambda} \qnorm{u_k^{\frac 1 2} P }_{2} \qnorm{ P u_j^{\frac 1 2} T_{j,k}}_{2}\le \pa{\max_{r \in \widetilde\Lambda}\qnorm{P u_r^{\frac 12}}_{2}} \sum_{k \in\widetilde\Lambda} \qnorm{ P u_j^{\frac 1 2} T_{j,k}}_{2},
\end{align}
and hence
\beq
\max_{r \in\widetilde\Lambda}\qnorm{P u_r^{\frac 12}}_{2}\le  \max_{j\in\widetilde\Lambda} { \sum_{k \in\widetilde\Lambda} \qnorm{ P u_j^{\frac 1 2} T_{j,k}}_{2}} .
\eeq

We have
\begin{align}\label{aftersa}
\qnorm{ P u_j^{\frac 1 2} T_{j,k}}_{2}^{2}& =\E\set{\tr \set{P u_j^{\frac 1 2} T_{j,k}T_{j,k}^{*}u_j^{\frac 1 2} P}}= \E\set{\tr \set{u_j^{\frac 1 2}P u_j^{\frac 1 2} T_{j,k}T_{j,k}^{*} }}\\
&   \le Q_{\mu_j}(\abs{I})\E_{\bom_{j}^{\perp}} \set{\tr T_{j,k}T_{j,k}^{*}}=Q_{\mu_j}(\abs{I}) \qnorm{ T_{j,k}}_{2}^2    ,\notag
\end{align}
where we used the basic spectral averaging estimate \eq{sa0}  (note that $ T_{j,k}$ does not depend on $\omega_{j}$).  It follows that
\begin{align} \label{maxEtrfull}
\max_{r \in\widetilde\Lambda} \E   \set{\tr P u_{r}}
 \le Q_{\Lambda}(\abs{I})  \pa{ \max_{j\in \widetilde\Lambda} \sum_{k \in\widetilde\Lambda}  \qnorm{ T_{j,k}}_{2}}^{2}.
 \end{align}

To prove (i), we use
 \eqref{qcomp} with
\begin{align}
\qnorm{ T_{j,k}}_{1}&\le u_-^{-1}\qnorm{\widetilde{P}_{j} u_{j}^{\frac 1 2}}_{2} \qnorm{\widetilde{P}_{j} u_{k}^{\frac 1 2}}_{2} \le   u_-^{-1} \max_{r \in\widetilde\Lambda}  \E   \set{\tr \widetilde{P}_{j} u_{r}} ,
\end{align}
to conclude that
\begin{align} \label{maxEtr}
\max_{r \in\widetilde\Lambda} \E   \set{\tr P u_{r}}
 \le  u_-^{-1}Q_{\Lambda}(\abs{I})\pa{\max_{j,k \in  \widetilde\Lambda}  \E   \set{\tr \widetilde{P}_{j} u_{k}} } \pa{ \max_{j\in \widetilde\Lambda} \sum_{k \in\widetilde\Lambda}  \qnorm{ T_{j,k}}_{\infty}^{\frac 12}}^{2}.
 \end{align}
 If the function $g$ in \eq{deftildeP} satisfies $g(E)=0$ for $E> E_1\ge E_0$,   it follows from the usual trace estimate
for Schr\" odinger operators (e.g., \cite[Lemma~A.4]{GKduke}) that
\beq\label{trPjuk3}
\tr \widetilde{P}_{j} u_{k} \le C_{d, \norm{ V_{\mathrm{per}}^-},\delta_+}(1 + E_1)^{2[[\frac d 4]]} \qtx{for all} j,k \in \widetilde\Lambda \qtx{and}  \bom \in [0,\infty[^{\Z^d},
\eeq
where $V_{\mathrm{per}}^-$ denotes the negative part of $V_{\mathrm{per}}$ and   $[[\frac d 4]]$ is the smallest integer $> \frac d 4$.  We  now take  $g(E)=g_0(E-E_0)$, where $g_0 \in C^\infty(\R)$, $0\le g_0\le 1$, $g_0(E)=1$ for $E\le 0$, and $g_0(E)=0$ for $E\ge 1$.  We now apply   \cite[Theorem~2]{GKdecay}, concluding that that for all $n\in \N$, $ j,k \in \widetilde\Lambda$, and  $\bom \in [0,\infty[^{\Z^d}$  we have
\beq\label{decayT1}
\norm{ T_{j,k}}\le u_-^{-\frac 12}\norm{u_j^{\frac 1 2}\widetilde{P}_j \chi_k}\le  u_-^{-\frac 12}\norm{\chi_{\Lambda_{\delta_+}(j) }\widetilde{P}_j \chi_{k }}\le
C_{d,\norm{ V_{\mathrm{per}}^-},\delta_+,n} u_-^{-\frac 1 2}\frac{1 +\log (1 + E_0)}{\pa{1+d_\Lambda(j,k)}^n},
\eeq
where $d_\Lambda(\ , \ )$ is the distance on the torus $\Lambda=\Lambda_L$:
\begin{equation} \label{distL}
{d_\Lambda}(y,y^\prime)= \min_{r\in L \Z^{d}} |y-y^\prime+r| \qtx{for} y,y^\prime\in {\Lambda}.
\end{equation}
(Note that  the results in  \cite{GKdecay} are valid on the torus with the appropriate modifications, the main one being the use of the distance on the torus.)
Taking $n=2d + 2$, and using
\beq
\sum_{k \in\widetilde\Lambda}\pa{1+d_\Lambda (j,k)}^{-(d+1)}\le \sum_{k \in\Z^d}\pa{1+\abs{k}}^{-(d+1)}< \infty \qtx{for all} j \in\widetilde\Lambda,
\eeq
we conclude that
\beq \label{sumTjk2}
 \pa{ \max_{j\in \widetilde\Lambda} \sum_{k \in\widetilde\Lambda}  \qnorm{ T_{j,k}}_{\infty}^{\frac 12}}^{2}\le  C_{d,\norm{ V_{\mathrm{per}}^-},\delta_+}  u_-^{-\frac 1 2}(1 +\log (1 + E_0)).
\eeq
It now follows from  \eq{maxEtr},  \eq{trPjuk3}, and \eq{sumTjk2} that
\begin{align} \label{maxEtr45}
\max_{r \in\widetilde\Lambda} \E   \set{\tr P u_{r}}
 \le    C_{d,\norm{ V_{\mathrm{per}}^-},\delta_+}(1 + E_0)^{2[[\frac d 4]]}(1 +\log (1 + E_0)) u_-^{-\frac 3 2}  Q_{\Lambda}(\abs{I}),
   \end{align}
which is  \eq{localWeg1}.  The Wegner estimate \eq{localWegreal} is an immediate consequence of \eq{localWeg1} and \eq{covLambda}.  This finishes the proof of (i).

Now  suppose that  the generalized Anderson Hamiltonian $H_\bom$ has a spine $\Gamma$ of order $K$.    For any $j \in \Z^d$ there exists a spine $\Gamma_j \subset \Gamma$ of order $2K$ with $j\notin \Gamma_j$, and we can write 
\beq\label{Homperp}
H_{\bom_j^\perp}= H_{{\bom}_{\Gamma_j}} +  V_{\bom_j^\perp \setminus{\bom}_{\Gamma_j}}, \qtx{where} 0\le   V_{\bom_j^\perp \setminus{\bom}_{\Gamma_j}}:= V_{\bom_j^\perp}-  V_{{\bom}_{\Gamma_j}}\le U_+.
\eeq
We take  the function $g$ in \eq{deftildeP} so $0\le g\le 1$, $g(E)=1$ for $E\le E_0$,  $g(E)=0$ for $E\ge E^*$; where  $E^*\ge E_0$ will be later  chosen appropriately. We have (writing $H_{{\bom}_j^{\perp}}$ for $H_{{\bom}_j^{\perp}}\up{\Lambda}$, etc.)
\begin{align}
\tr \set{\widetilde{P}_{j} u_k} 
& \le \e^{tE^*}\tr \set{ \e^{- t H_{{\bom}_j^{\perp}}  } u_k} \le \e^{tE^*} \tr \set{\e^{- t H_{\bom_{\Gamma_{j}}}}u_k}\qtx{for} t>0, \label{usepositivity}
\end{align}
 where we used \eq{Homperp} and  the positivity preserving property  as in \cite[Lemma~2.2]{BGKS}.
Setting 
\beq
P_{\bom_{\Gamma_{j}}}([0,E]):= \chi_{[0,E]}(H_{\bom_{\Gamma_{j}}}\up{\Lambda})=\chi_{]-\infty,E]}(H_{\bom_{\Gamma_{j}}}\up{\Lambda}),
\eeq
we get, again using the positivity preserving property as in \cite[Lemma~2.2]{BGKS}, and requiring $t\ge 2$, that for all $E>0$ we have 
\begin{align}
\notag
 \tr \set{ \e^{-  t H_{\bom_{\Gamma_{j}}}}u_{k}}&\le   \tr \set{ P_{\bom_{\Gamma_{j}}}([0,E])u_{k}} + \e^{- \frac t 2 E }\tr \set{ \e^{- \frac t 2 H_{\bom_{\Gamma_{j}}}}u_{k}}\\  
  \label{boundsublatA}
 &\le  \tr \set{ P_{\bom_{\Gamma_{j}}}([0,E])u_{k}} +  \e^{- \frac t 2 E }\tr \set{ \e^{-   \frac t 2 H_{0}}u_{k}}\\
 &\le  \tr \set{ P_{\bom_{\Gamma_{j}}}([0,E])u_{k}}+  \e^{- \frac t 2 E }\tr \set{ \e^{-  H_{0}} u_{k}} \notag \\
 &\le   \tr \set{ P_{\bom_{\Gamma_{j}}}([0,E])u_{k}}+
  C_{d,V_{\mathrm{per}}, \delta_+}  \e^{- \frac t 2 E } .  \notag
 \end{align}
  Since $\Gamma_{j}$ is a spine of order $2K$ and $L\in 2qK\N$, the random operator 
$H_{\bom_{\Gamma_{j}}}\up{\Lambda}$ is covariant in the torus $\Lambda$,  and we have
 \begin{align}
 \E  \set{\tr \set{ P_{\bom_{\Gamma_{j}}}([0,E])u_{k}}} & = \frac 1{\#(\Gamma_{j}\cap\Lambda)} \sum_{r \in k+ \Gamma_{j}\cap\Lambda} \E_{\bom_{\Gamma_{j}}}\set{\tr \set{ P_{\bom_{\Gamma_{j}}}([0,E])u_{r}}}
 \notag \\
 & \le \frac {(2K)^d} {\abs{\Lambda}} U_+\E_{{\bom}_{\Gamma_j}}\set{ \tr { P_{\bom_{\Gamma_{j}}}([0,E])}} \label{translA}
 \end{align}
for all $E>0$.  Combining \eq{usepositivity},  \eq{boundsublatA}, and \eq{translA}  we get
  \beq   \label{boundsublatAB}
 \E\set{\tr \set{\widetilde{P}_{j} u_k} }\le   C_{d,V_{\mathrm{per}}, \delta_+,K}\, \e^{tE^*} \pa{ \abs{\Lambda}^{-1}\E_{{\bom}_{\Gamma_j}}\set{ \tr { P_{\bom_{\Gamma_{j}}}([0,E])}} +  \e^{- \frac t 2 E }}.
 \eeq

To prove  (ii), we take $E_0 \in ]0,\frac 18]$, fix $E^*=2E_0$, and    require $g \in C^\infty(\R)$ with   $\abs{g^{(j)}(E)}\le C E_0^{-j}$ for all $E\in \R$ and $j=1,2,\ldots,2d+4$, where $C$ is a constant independent of $E$.   Appplying    \cite[Theorem~2]{GKdecay} as in \eq{decayT1}, we get
\beq\label{decayT123}
\norm{ T_{j,k}}\le 
C_{d,\norm{ V_{\mathrm{per}}^-},\delta_+} u_-^{-\frac 1 2}E_0^{-2d -3 }{\pa{1+d_\Lambda(j,k)}^{-2d-2}},
\eeq
and conclude, similarly to \eq{sumTjk2}
\beq \label{sumTjk245}
 \pa{ \max_{j\in \widetilde\Lambda} \sum_{k \in\widetilde\Lambda}  \qnorm{ T_{j,k}}_{\infty}^{\frac 12}}^{2}\le  C_{d,\norm{ V_{\mathrm{per}}^-},\delta_+}  u_-^{-\frac 1 2}E_0^{-2d -3 }.
\eeq
Thus, it follows from  \eq{maxEtr} and \eq{sumTjk245} that
\begin{align} \label{maxEtr4578}
\max_{r \in\widetilde\Lambda} \E   \set{\tr P u_{r}}
 \le    C_{d,\norm{ V_{\mathrm{per}}^-},\delta_+} u_-^{-\frac 3 2} E_0^{-2d -3 } Q_{\Lambda}(\abs{I})\pa{\max_{j,k \in  \widetilde\Lambda}  \E   \set{\tr \widetilde{P}_{j} u_{k}} }.
   \end{align}

Note that $H_{{\bom}_{\Gamma_j}}$   would be an Anderson Hamiltonian but for the fact that the random potential is located on  $\Gamma_j$ instead of  $\Z^d$. All the results for Anderson Hamiltonians apply to $H_{{\bom}_{\Gamma_j}}$,  with the obvious modifications. $H_{{\bom}_{\Gamma_j}}$  is a  $2qK\Z^d$-ergodic  family of random self-adjoint operators. It has an integrated density of states $N_{\Gamma_j}(E)$, defined similarly to \eq{N(E)}, a continuous function in view of the Wegner estimate \eq{Wegnocov}.
It follows from \eqref{N(E)} that for all  $E\in\R$ there exists $L(E)$ such that for all boxes $\Lambda=\Lambda_L$ with $L \ge L(E)$ we have 
\beq  \label{Nineq}
\abs{\Lambda}^{-1} \E \pa{\tr  \, \chi_{]- \infty,E]}\pa{H_{{\bom}_{\Gamma_j}}\up{\Lambda}}}\le 2 N_{\Gamma_j}(E)  .
\eeq
 $N_{\Gamma_j}(E)$
  satisfies the  Lifshitz tails  estimate  \eq{LT}, so it follows that  given $\eta\in ]0, \frac 12]$ there exists $E^\ast(\eta)>0$ such  that 
\begin{equation}\label{NLiftail}
N_{\Gamma_j}\le  \e^{- E^{- \frac d2 +\eta} } \qtx{for all}  E \in [0, E^\ast(\eta)] .
\end{equation}
We conclude that
\beq
\abs{\Lambda_L}^{-1} \E \pa{\tr  \, \chi_{]- \infty,E]}\pa{H_{{\bom}_{\Gamma_j}}\up{\Lambda_L}}} \le 2  \e^{- E^{- \frac d2 +\eta} } \; \text{for}\; E \in [0, E^\ast(\eta)], \; L \ge L(E). \label{estNvol}
\eeq
In particular, requiring $8 E_0 \le E^\ast(\eta)$ and $L \ge L(8E_0)$, it follows from \eqref{boundsublatAB} with $E^*=2E_0$ and $E=8E_0$, \eq{Nineq}, and \eq{NLiftail},   that 
 \beq   \label{boundsublat294}
 \E\set{\tr \set{\widetilde{P}_{j} u_k} }\le   C_{d,V_{\mathrm{per}}, \delta_+,K}\,  \e^{2t E_0 } \pa{  \e^{- (8E_0)^{- \frac d2 +\eta} }+  \e^{- 4t E_0 }}.
 \eeq
We now choose  $t$ by ($t \ge 2$ since $E_0\le \frac 1 8$)
\beq
 \e^{- (8E_0)^{- \frac d2 +\eta} }= \e^{- 4t E_0 }, \qtx{i.e.,} t= \tfrac 1 4 \pa{8E_0}^{-1-  \frac d2 +\eta},
\eeq
getting
\begin{align}\label{Ptildeest1}
\E\set{\tr \set{\widetilde{P}_{j} u_k} }&\le  2 C_{d,V_{\mathrm{per}}, \delta_+,K} \,  \e^{2t E_0 }  \e^{- 4t E_0 }=   2 C_{d,V_{\mathrm{per}}, \delta_+,K} \, \e^{-2t E_0 }  \\
&=  2 C_{d,V_{\mathrm{per}}, \delta_+,K} \,   \e^{- \frac 1 2(8E_0)^{- \frac d2 +\eta} }. \notag
\end{align}

Thus, if  $8 E_0 \le E^\ast(\eta)$ and $L \ge L(8E_0)$ it follows from\eq{maxEtr4578} and \eq{Ptildeest1} that
\begin{align} \label{maxEtr45784}
\max_{r \in\widetilde\Lambda} \E   \set{\tr P u_{r}}
 \le    C_{d, V_{\mathrm{per}},\delta_+,K} \, u_-^{-\frac 3 2}  Q_{\Lambda}(\abs{I}) E_0^{-2d -3 } \e^{- \frac 1 2(8E_0)^{- \frac d2 +\eta} }.
   \end{align}
It follows that there is $E^\ddagger(\eta)=E^\ddagger(\eta,d, V_{\mathrm{per}},\delta_+,u_-,K,\mu_\Gamma)>0$ such that for  $ E_0 \le  E^\ddagger(\eta)$ and $L \ge L(8E_0)$ we get
\begin{align} \label{maxEtr45789}
\max_{r \in\widetilde\Lambda} \E   \set{\tr P u_{r}}
 \le  \e^{-E_0^{- \frac d2 +\eta} }   Q_{\Lambda}(\abs{I}) \qtx{for} I \subset [0,E_0]  ,
   \end{align}
which is \eq{localWeg344}. Thus   (ii) is proven.

To prove (iii), we also assume that  $\mu_\Gamma$ is H\" older continuous, so
   \eq{Wegnocov} and \eq{WegnocovM} yield a Wegner estimate that allows the performance of the bootstrap multiscale analysis  \cite{GKboot,Kle} for the random Schr\"odinger operator $H_{{\bom}_{\Gamma_j}}$,
and hence for $H_{\bom_j^\perp}$ by treating $V_{\bom_j^\perp \setminus{\bom}_{\Gamma_j}}$ in \eq{Homperp} as a fixed nonnegative uniformly  bounded   background potential as in \cite{GKjems}.   The `a priori' finite volume estimate required for starting the multiscale analysis is given by \cite[Proposition~4.3]{GKjems}.  It follows that there exists $E_1>0$ such that we can perform a bootstrap  multiscale analysis  for $H_{\bom_j^\perp}$ (using only the random variables ${\bom}_{\Gamma_j}$), the constants being uniform in $j\in \Z^d$.  In particular, taking $0< E_0\le E_1$, $g=\chi_{]-\infty,E_0]}$ (in particular, $E^*=E_0$), so $\widetilde{P}_j= \chi_{]-\infty,E_0]}(H_{\bom_j^\perp})$,  we conclude that
(this follows from the multiscale analysis as in \cite{GKboot,Kle}, see also \cite{R}; the argument holds in finite volume) for $L$ large (how large depending on $d, V_{\mathrm{per}},\delta_\pm,u_-,\mu_\Gamma,K$)
\beq\label{decayT22}
\qnorm{ T_{j,k}}_2\le u_-^{-\frac 1 2}\norm{\chi_{\Lambda_{\delta_+}(j) }\widetilde{P}_j {\chi}_k}_2\le
C_{d, V_{\mathrm{per}},\delta_\pm,u_-} \e^{- \sqrt{d_\Lambda(j,k)}} \qtx{for} j,k\in  \widetilde\Lambda.
\eeq
In particular,  give  $s>0$,  we have 
\beq \label{decayT22s}
\sum_{k \in\widetilde\Lambda}\e^{-s \sqrt{d_\Lambda(j,k)}}\le \sum_{k \in\Z^d}\e^{-s\abs{k}^{\frac 1 2}}= C_{d,s} < \infty \qtx{for all} j \in\widetilde\Lambda.
\eeq

Since we also  have
\beq\label{TjkTj}
\qnorm{ T_{j,k}}_2=\qnorm{u_j^{\frac 1 2}\widetilde{P}_j \hat{\chi}_k}_2 \le u_-^{-\frac 1 2}\qnorm{u_j^{\frac 1 2}\widetilde{P}_j }_2= u_-^{-\frac 1 2} \pa{ \E \set{\tr \widetilde{P}_j u_j}}^{\frac 1 2},
\eeq
it follows from \eq{maxEtrfull}, \eq{decayT22}, \eq{decayT22s}, and \eq{TjkTj}, that for any $\eta \in ]0,1[$ we have
\beq \label{maxEtrfull56}
\max_{r \in\widetilde\Lambda} \E   \set{\tr P u_{r}}\le C_{d, V_{\mathrm{per}},\delta_\pm,u_-,\eta} \, Q_{\Lambda}(\abs{I}) \pa{\max_{j \in\widetilde\Lambda} \E \set{\tr \widetilde{P}_j u_j}}^{1-\eta}.
\eeq

We now consider  energies   $0<  E_2\le E_3$; we will fix $E_3$ later. 
 It follows from  \eq{boundsublatAB} with $E^*=E_0$ ,  $E=E_2$, and $t=\frac 1 {E_0}$, that 
  \beq   \label{boundsublatABC}
 \E\set{\tr \set{\widetilde{P}_{j} u_k} }\le   C_{d,V_{\mathrm{per}}, \delta_+,K} \pa{ \abs{\Lambda}^{-1}\E_{{\bom}_{\Gamma_j}}\set{ \tr { P_{\bom_{\Gamma_{j}}}([0,E_2])}} +  \e^{- \frac  {E_{2}}{2E_0}}}.
 \eeq
Using   \eq{Wegnocov} and \eq{Holderctsm}, we get 
 \begin{align}  \label{transl} 
 \E  \set{\tr \set{ P_{\bom_{\Gamma_{j}}}([0,E_2]) }} \le C_{E_{3}} Q\up{m_d}_{\Gamma}(E_{2})
 \le C_{E_{3}}  C_{\mu_\Gamma,\alpha,m_d}E_2^\alpha ,
 \end{align}
 the constant $C_{E_{3}}$ depending only on $d, V_{\mathrm{per}},\delta_+,u_\pm,K$ and on  $ E_3$.  Combining \eq{boundsublatABC} and \eq{transl}  we get
  \beq\label{trPQm}
 \E\set{\tr \set{\widetilde{P}_{j} u_j} }\le C_{1} \pa{ C_{\mu_\Gamma}E_2^\alpha +  \e^{- \frac  {E_{2}}{2 E_{0}}} },
 \eeq
 with a constant $C_{1}=C_{d, V_{\mathrm{per}},\delta_+,u_\pm, K,E_{1},E_3}$ and $C_{\mu_\Gamma}=  C_{\mu_\Gamma,\alpha,m_d}$.

Let $\beta(s)$ be defined   on $[0,\infty[$  by $\beta (0)=0$ and 
\beq \label{defbeta}
C_{\mu_\Gamma} \pa{\beta(s)}^\alpha = \e^{- \frac {\beta(s)}{2 s}} \qtx{for} s >  0.
\eeq
In particular,
\beq
C_{\mu_\Gamma}\pa{\beta(s)}^\alpha \e^{\frac {\beta(s)}{2 s}} =1 , \qtx{i.e.,}
\tfrac {\beta(s)}{2 \alpha s} \e^{\frac {\beta(s)}{2\alpha s}} = \pa{2\alpha s C_{\mu_\Gamma}^{\frac 1 \alpha}}^{-1}.
\eeq
If  
 \beq
  \pa{2\alpha s C_{\mu_\Gamma}^{\frac 1 \alpha}}^{-1}\ge 3, \qtx{i.e.,} 6\alpha s C_{\mu_\Gamma}^{\frac 1 \alpha} \le 1,
 \eeq
 we have
 \beq\label{betabbd}
 \tfrac {\beta(s)}{2 \alpha s} \le \log \pa{2\alpha s C_{\mu_\Gamma}^{\frac 1 \alpha}}^{-1}, \qtx{i.e.,} \beta(s) \le 2 \alpha s  \log \pa{2\alpha s C_{\mu_\Gamma}^{\frac 1 \alpha}}^{-1}.
 \eeq

We now choose $E_2= \beta(E_0)$ and  $E_3=\beta(E_1)$, and require
\beq
E_0 \le \widetilde{E}_1= \min \set{E_1,  \pa{6\alpha C_{\mu_\Gamma}^{\frac 1 \alpha}}^{-1}}.
\eeq
It follows from \eq{trPQm}  and \eq{betabbd}  that
 \begin{align}\label{trPubeta}
\E\set{\tr \set{\widetilde{P}_{j} u_j} }\le 2C_1 C_{\mu_\Gamma} \pa{ 2 \alpha E_0  \log \tfrac 1{2\alpha E_0 C_{\mu_\Gamma}^{\frac 1 \alpha}}}^\alpha.  
\end{align}

The estimate \eq{localWeg3} follows immediately from \eq{maxEtrfull56} and \eq{trPubeta}.  This proves (iii).
\end{proof}


\end{document}